\documentclass[twocolumn,showpacs,preprintnumbers,amsmath,amssymb]{revtex4}

\usepackage{graphicx}
\usepackage{dcolumn}
\usepackage{bm}

\renewcommand{\d}{{\rm d}}
\newcommand{\bbbone}{\mathchoice {\rm 1\mskip-4mu l} {\rm 1\mskip-4mu l}
{\rm 1\mskip-4.5mu l} {\rm 1\mskip-5mu l}}

\newcommand{\scalprod}[2]{\left\langle {#1}, {#2}\right\rangle}
\newcommand{\fer}[1]{(\ref{#1})}
\newcommand{\av}[1]{\left\langle{#1}\right\rangle}

\newcommand{\s}{{\rm S}}
\newcommand{\h}{{\cal H}}
\newcommand{\hh}{{\frak h}}
\newcommand{\cx}{{\mathbb C}}
\newcommand{\rhobar}{{\overline \rho}}
\newcommand{\e}{{\,\rm e}}
\renewcommand{\i}{{\rm i}}
\newcommand{\rx}{{\mathbb R}}
\newcommand{\tr}{{\rm Tr}}
\renewcommand{\r}{{\rm R}}

\begin{document}


\title{Decoherence and Thermalization}

\author{M. Merkli}
\email{merkli@math.mun.ca}
 \altaffiliation[Present address: ]{Department of Mathematics and Statistics,
Memorial University of Newfoundland,
St. John's, NL,
Canada A1C 5S7}
 \altaffiliation[Supported by NSERC under grant NA 7901.]{}
\author{I.M. Sigal}
\email{im.sigal@utoronto.ca}
\altaffiliation[Supported by NSERC under grant NA 7901.]{}
\affiliation{Department of Mathematics, University of Toronto, Toronto, Ontario, Canada M5S 2E4}
\author{G.P. Berman}
\email{gpb@lanl.gov}
\altaffiliation[Supported by the NNSA of the U.S. DOE at LANL under Contract No. DE-AC52-06NA25396.]{}
\affiliation{Theoretical Division, MS B213, Los Alamos National Laboratory, Los Alamos, NM 87545, USA}

\date{\today}

\begin{abstract}
We present a rigorous analysis of the phenomenon of decoherence for general $N$-level systems coupled to reservoirs of free massless bosonic fields. We apply our general results to the specific case of the qubit. Our approach does not involve master equation approximations and applies to a wide variety of systems which are not explicitly solvable.
\end{abstract}

\maketitle


\section{\label{sect:introduction} Introduction}

We examine rigorously the phenomenon of quantum decoherence. This phenomenon is brought about
by the interaction of a quantum system, called in what follows  ``the system $\s$'', with an environment, or ``reservoir $\r$''.
Decoherence is reflected in the temporal decay of off-diagonal elements of the reduced density matrix of the
system in a given basis. The latter is determined by the measurement to be performed. To our knowledge, this phenomenon has been analyzed rigorously so far only for explicitly solvable
models, see e.g. \cite{DBV,DG,JZKGKS,MP,SGC,PSE,VK}. In this paper we consider the decoherence phenomenon for quite
general non-solvable models. Our analysis is based on the modern theory of resonances for quantum statistical
systems as developed in \cite{BFS0,JP1,BFS,JP2,HuS,SV,MMS1,MMS2} (see also the book \cite{HiS}), which is
related to resonance theory in non-relativistic quantum electrodynamics \cite{BFS,BCFS}.

Let $\hh=\hh_\s\otimes\hh_\r$ be the Hilbert space of the system
interacting with the environment, and let
\begin{equation}
H= H_\s\otimes\bbbone_\r + \bbbone_\s\otimes H_\r +\lambda v
\label{ii1}
\end{equation}
be its Hamiltonian. Here, $H_\s$ and $H_\r$ are the Hamiltonians of the system and the reservoir, respectively, and $\lambda v$ is an interaction with a coupling constant $\lambda\in\rx$. In the following we will omit trivial factors $\bbbone_\s\otimes$ and  $\otimes\bbbone_\r$. The reservoir is taken initially in an equilibrium state at some temperature $T=1/\beta>0$. Let $\rho_t$ be the density matrix of the total system at time $t$. The reduced density matrix (of the system $\s$) at time $t$ is then formally given by
\begin{equation}
\rhobar_t = \tr_\r\,\rho_t,
\label{rdm}
\end{equation}
where $\tr_\r$ is the partial trace with respect to the reservoir degrees of freedom. Formulas \fer{ii1} and \fer{rdm} describe the situation where a state of the reservoir is given by a well-defined density matrix on the Hilbert space $\hh_\r$. In order to describe decoherence and thermalization we need to consider ``true'' (dispersive) reservoirs, obtained for instance by taking a thermodynamic limit, or a continuous-mode limit. We refer to \cite{MSB2} for a detailed description of such reservoirs, which is not needed in the presentation of our results here.

Let $\rho(\beta,\lambda)$ be the equilibrium state of the interacting system at temperature $T=1/\beta$ and set $\rhobar(\beta,\lambda):=\tr_\r\rho(\beta,\lambda)$. There are three possible scenarios for the asymptotic behaviour of the reduced density matrix as $t\rightarrow\infty$:
\begin{itemize}
\item[(i)]  $\rhobar_t\longrightarrow \rhobar_\infty = \rhobar(\beta,\lambda)$,
\item[(ii)] $\rhobar_t\longrightarrow \rhobar_\infty \neq  \rhobar(\beta,\lambda)$,
\item[(iii)] $\rhobar_t$ does not converge.
\end{itemize}
The first situation is generic while the last two are not, although they are of interest, e.g. for energy conserving, or quantum non-demolition interactions characterized by $[H_\s,v]=0$, see \cite{JZKGKS, MSB2}.

Decoherence is a basis-dependent notion. It is usually defined as the vanishing of the off-diagonal elements $[\rhobar_t]_{m,n}$, $m\neq n$ in the limit $t\rightarrow\infty$, in a chosen basis. Most often decoherence is defined w.r.t. the basis of eigenvectors of the system Hamiltonian $H_\s$ (the energy, or computational basis for a quantum register), though other bases, such as the position basis for a particle in a scattering medium \cite{JZKGKS}, are also used.

Since $\rhobar(\beta,\lambda)$ is generically non-diagonal in the energy basis, the off-diagonal elements of $\rhobar_t$ will not vanish in the generic case, as $t\rightarrow\infty$. Thus, strictly speaking, decoherence in this case should be defined as the decay (convergence) of the off-diagonals of $\rhobar_t$ to the corresponding off-diagonals of $\rhobar(\beta,\lambda)$. The latter are $O(\lambda)$.
If these terms are neglected then decoherence manifests itself as a process in which initially coherent superpositions of basis elements $\psi_j$ become incoherent statistical mixtures,
$$
\sum_{j,k}c_{j,k}|\psi_j\rangle\langle\psi_k| \longrightarrow \sum_j p_j |\psi_j\rangle\langle\psi_j|,\ \ \ \mbox{ as $t\rightarrow\infty$}.
$$
In particular, phase relations encoded in the $c_{j,k}$ disappear
for large times.

\section{\label{sect:results} General Results}

We consider $N$-dimensional quantum systems interacting with reservoirs of massless free quantum fields (photons, phonons or other massless excitations) through an interaction $v=G\otimes \varphi(g)$, see also \fer{ii1} and \fer{michael*}. Here, $G$ is a hermitian $N\times N$ matrix and $\varphi(g)$ is the bosonic field operator smoothed out with the form factor $g(k)$, $k\in{\mathbb R}^3$. For any observable $A$ of the system we set
\begin{equation}
\av{A}_t := \tr_\s(\rhobar_t A) =\tr_{\s +\r}(\rho_t (A\otimes\bbbone_\r) ).
\label{avA}
\end{equation}
Assuming certain regularity conditions on $g(k)$ (allowing e.g. $g(k)= |k|^p\e^{-|k|^m}g_1(\sigma)$ where $g_1$ is a function on the sphere and where $p=-1/2+n$, $n=0,1,\ldots$, $m=1,2$), we show that the ergodic averages
\begin{equation*}
\av{\av{A}}_\infty := \lim_{T\rightarrow\infty}\frac 1T\int_0^T\av{A}_t \d t
\end{equation*}
exist, i.e., that $\av{A}_t$ converges in the ergodic sense as $t\rightarrow\infty$. Furthermore, we show that for $t\geq 0$, and for any $0<\tau'<\frac{2\pi}{\beta}$,
\begin{equation}
\av{A}_t-\av{\av{A}}_\infty =\sum_{\varepsilon \neq 0}\e^{\i t\varepsilon} R_\varepsilon(A) +O(\lambda^2\e^{-\tau' t/2}),
\label{introdd1}
\end{equation}
where the complex numbers $\varepsilon$ are the  eigenvalues of a certain explicitly given operator $K(\tau')$, lying in the strip $\{z\in\cx\ |\ 0\leq {\mathrm{Im}} z<\tau'/2\}$.  They have the expansions
\begin{equation}
\varepsilon\equiv \varepsilon_e^{(s)} = e -\lambda^2\delta_e^{(s)} + O(\lambda^4),
\label{int1}
\end{equation}
where $e\in\mathrm{spec}(H_\s\otimes\bbbone_\s-\bbbone_\s\otimes
H_\s) = \mathrm{spec}(H_\s)- \mathrm{spec}(H_\s)$ and the 
$\delta_e^{(s)}$ are the eigenvalues of a matrix $\Lambda_e$, called
a {\it level-shift operator}, acting on the eigenspace of
$H_\s\otimes\bbbone_\s-\bbbone_\s\otimes H_\s$ corresponding to the
eigenvalue $e$ (which is a subspace of $\hh_\s\otimes\hh_\s$). The
level shift operators play a central role in the ergodic theory of
open quantum systems, see e.g. \cite{Mlso, MSB2}. 

The coefficients $R_\varepsilon(A)$ in
\fer{introdd1} are linear functionals of $A$ which depend on the initial state $\rho_0$ and the Hamiltonian $H$. They have the expansion
$
R_\varepsilon(A)=\sum_{(m,n)\in I_e}\varkappa_{m,n} A_{m,n} +O(\lambda^2),
$
where $I_e$ is the collection of all pairs of indices such that $e=E_m-E_n$, the $E_k$ being the eigenvalues of $H_\s$. Here, $A_{m,n}$ is the $(m,n)$-matrix element of the observable $A$ in the energy basis of $H_\s$, and the $\varkappa_{m,n}$ are coefficients depending on the initial state of the system (and on $e$, but not on $A$ nor on $\lambda$).

\section{\label{sect:qubit} Qubit}

Our results for the qubit can be summarized as follows. Consider a linear coupling,
\begin{equation}
v = \left[
\begin{array}{cc}
 a & c\\
\overline c & b
\end{array}
\right]
\otimes\varphi(g),
\label{michael*}
\end{equation}
where $\varphi(g)$ is the Bose field operator as above. The form-factor $g\in L^2({\mathbb R}^3,\d^3k)$ contains an ultra-violet cutoff which
introduces a time-scale $\tau_{UV}$. This time scale depends on the
physical system in question. We can think of it as coming from some frequency-cutoff determined by a characteristic length scale beyond which the interaction decreases rapidly. For instance, for a phonon field $\tau_{UV}$ is naturally identified with the inverse of the Debye frequency. We assume $\tau_{UV}$ to be much smaller than the time scales considered here.

A key role in the decoherence analysis is played by the infrared
behaviour of form factors $g(k)$. We characterize
this behaviour by the unique $p\geq -1/2$ satisfying
\begin{equation}
0< \lim_{|k|\rightarrow 0} \frac{|g(k)|}{|k|^p}=C<\infty. 
\label{38}
\end{equation}
The power $p$ depends on the physical model considered, e.g.
for quantum-optical systems $p=1/2$. We can treat $p=-1/2+n$, $n=0,1,\ldots$. 

Decoherence in models with interaction \fer{michael*} with $c=0$ is
considered in \cite{BBD,BBBD,DBV,DG,JZKGKS,MP,PSE,SGC,U,MSB2}. This is the  situation of a non-demolition
(energy conserving) interaction, where $[v,H_\s]=0$, and consequently energy-exchange processes are
suppressed. The resulting decoherence is called phase-decoherence. A
particular model of phase-decoherence is obtained by the so-called
position-position coupling, where the matrix in the interaction
\fer{michael*} is the Pauli matrix $\sigma_z$ \cite{BBBD,DG,PSE,U}.
On the other hand,
energy-exchange processes, responsible for driving the system to
equilibrium, have a probability proportional to $|c|^{2n}$, for some
$n\geq 1$ (and $a$, $b$ do not enter)
\cite{BFS,FM1,JP1,Mlso,MMS1,MMS2}. Thus the property $c\neq 0$ is
important for thermalization (return to equilibrium).

We express the energy-exchange effectiveness by the function
\begin{equation*}
\xi(\eta) = \lim_{\epsilon\downarrow 0} \frac 1\pi\int_{\rx^3}\d^3k\coth\!\left(\frac{\beta |k|}{2}\right)
 |g(k)|^2 \frac{\epsilon}{(|k|-\eta)^2+\epsilon^2},
\end{equation*}
where $\eta\geq 0$ represents the energy at which processes between
the qubit and the reservoir take place. Let $\Delta=E_2-E_1>0$ be
the energy gap of the qubit. In works on convergence to equilibrium
it is usually assumed that $|c|^2\xi(\Delta)>0$. This condition is
called the ``Fermi Golden Rule Condition''. It means that the
interaction induces second-order ($\lambda^2$) energy exchange
processes at the Bohr frequency of the qubit (emission and
absorption of reservoir quanta). The condition $c\neq 0$ is actually
{\it necessary} for thermalization while $\xi(\Delta)>0$ is not
(higher order processes can drive the system to equilibrium).
Observe that $\xi(\Delta)$ converges to a fixed function, as $T\rightarrow 0$, and increases exponentially as
$T\rightarrow \infty$. The expression for decoherence times involves
also $\xi(0)$, see \fer{i4}.

Our analysis allows to describe the dynamics of systems which exhibit {\it both} thermalization {\it and} (phase) decoherence. 

Let the initial density matrix, $\rho_{t=0}$, be of the form
$\rhobar_0\otimes\rho_{\r,\beta}$. (Our method does not require the
initial state to be a product, see \cite{MSB2}.) Denote by
$p_{m,n}$ the operator represented in the energy
basis by the $2\times 2$ matrix whose entries are zero, except the
$(n,m)$ entry which is one. We show that for $t\geq 0$
\begin{eqnarray}
&&[\rhobar_t]_{1,1} -\av{\av{p_{1,1}}}_\infty = \e^{\i
t\varepsilon_0(\lambda)} \left[ C_0
+O(\lambda^2)\right]\label{i1}\\
&&+ \e^{i t\varepsilon_\Delta(\lambda)} O(\lambda^2) + \e^{i t\varepsilon_{-\Delta}(\lambda)} O(\lambda^2)
+ O(\lambda^2\e^{-t\tau'/2})
\nonumber
\end{eqnarray}
and
\begin{eqnarray}
&&[\rhobar_t]_{1,2} -\av{\av{p_{2,1}}}_\infty = \e^{\i
t\varepsilon_\Delta(\lambda)} \left[C_\Delta
+O(\lambda^2)\right]\label{i2}\\
&&+ \e^{i t\varepsilon_0(\lambda)} O(\lambda^2) + \e^{i t\varepsilon_{-\Delta}(\lambda)} O(\lambda^2)
+ O(\lambda^2\e^{-t\tau'/2}).
\nonumber
\end{eqnarray}
Here, $C_0, C_\Delta$ are explicit constants depending on the initial condition $\rhobar_0$, but not on $\lambda$, and the resonance energies $\varepsilon$ have the expansions
\begin{eqnarray}
\varepsilon_0(\lambda) &=& \i\lambda^2 \pi^2|c|^2\xi(\Delta) +O(\lambda^4)\nonumber\\
\varepsilon_\Delta(\lambda) &=& \Delta +\lambda^2 R +{\textstyle \frac \i2}\lambda^2\pi^2\left[ |c|^2 \xi
(\Delta)+(b-a)^2\xi(0)\right]\nonumber\\
&&  +O(\lambda^4)\label{i4}
\end{eqnarray}
and $\varepsilon_{-\Delta}(\lambda)= -\overline{\varepsilon_\Delta(\lambda)}$, with the real number
\begin{eqnarray*}
\lefteqn{R = {\textstyle \frac 12}(b^2-a^2) \scalprod{g}{\omega^{-1}g}}\\
&& +{\textstyle \frac 12}|c|^2 {\rm P.V.}\int_{\rx\times S^2} u^2|g(|u|,\sigma)|^2 \coth\!\left(\frac{\beta |u|}{2}\right)\frac{1}{u-\Delta}.
\end{eqnarray*}
The error terms in \fer{i1}, \fer{i2} and \fer{i4} satisfy (for
small $\lambda$):
$
\left|\frac{O(\lambda^2)}{\lambda^{2}}\right|<C \mbox{\ \ and\ \ }
\sup_{t\geq
0}\left|\frac{O(\lambda^2\e^{-t\tau'/2})}{\lambda^{2}\e^{-t\tau'/2}}\right|<C.
$

To our knowledge this is the first time that formulas for the decay
of off-diagonal matrix elements of the reduced density matrix are
obtained for models which are not explicitly solvable, and without using
uncontrolled master equation approximations (see e.g. \cite{BBBD} and references therein).

{\it Remarks.\ } 1)\ The corresponding expressions for the matrix elements
$[\rhobar_t]_{2,2}$ and $[\rhobar_t]_{2,1}$ are obtained from the
relations $[\rhobar_t]_{2,2}=1-[\rhobar_t]_{1,1}$ (conservation of
unit trace) and $[\rhobar_t]_{2,1} = [\rhobar_t]_{1,2}^*$
(hermiticity of $\rhobar_t$).

2)\ If the qubit is initially in one of the logic pure states
$\rhobar_0=|\varphi_j\rangle\langle\varphi_j|$, where
$H_\s\varphi_j=E_j\varphi_j$, $j=1,2$, then  we find $C_\Delta=0$,
and $C_0=\e^{\beta\Delta/2}(\e^{\beta\Delta}+1)^{-3/2}$ for $j=1$
and $C_0=\e^{\beta\Delta}(\e^{\beta\Delta}+1)^{-3/2}$ for $j=2$, see \cite{MSB2}.

3)\ To second order in $\lambda$, the imaginary part of
$\varepsilon_\Delta$  is increased by a term $\propto (b-a)^2\xi(0)$
only if $p=-1/2$, where $p$ is defined in \fer{38}. For $p>-1/2$ we
have $\xi(0)=0$ and that contribution vanishes. For $p<-1/2$ we have
$\xi(0)=\infty$.

4)\ It is easy to see that $\xi(\Delta)$ and $R$ contain purely
quantum, vacuum fluctuation terms as well as thermal ones, while $\xi(0)$
is determined entirely by thermal fluctuations; it is proportional
to $\beta^{-1}= T$.

5)\ The second order difference $D$, defined by $\mathrm{Im}\varepsilon_0(\lambda) -\mathrm{Im}\varepsilon_\Delta(\lambda)=
\lambda^2 D+O(\lambda^4)$, is
$
D={\textstyle \frac 12}\pi^2\left[ |c|^2\xi(\Delta) -(b-a)^2\xi(0)\right].
$
For $D>0$ the populations converge to their limiting values faster than the off-diagonal matrix elements, as $t\rightarrow\infty$ (coherence persists beyond thermalization of the populations). For $D<0$ the off-diagonal elements converge faster. If the interaction matrix is diagonal ($c=0$) then $D\leq 0$, if it is off-diagonal then $D\geq 0$.

6)\ For energy-conserving interactions, $c=0$, it follows that full
decoherence occurs if and only if $b\neq a$ and $\xi(0)>0$. If
either of these conditions are not satisfied then the off-diagonal
matrix elements are purely oscillatory (while the populations are
constant), see also \cite{MSB2}. 

{\it Illustration.\ } Let the initial state of $\s$ be given by a coherent superposition in the energy basis,
\begin{equation}
\rhobar_0=\textstyle \frac 12
\left[
\begin{array}{cc}
 1 & 1\\
 1 & 1
\end{array}
\right].
\label{illust}
\end{equation}
We obtain the following expressions for the dynamics of the reduced matrix elements,
for all $t\geq 0$:
\begin{eqnarray*}
[\rhobar_t]_{m,m} &=& \frac{\e^{-\beta E_m}}{Z_{\s,\beta}} +\frac{(-1)^m}{2} \tanh\left(\frac{\beta\Delta}{2}\right) \e^{\i t\varepsilon_0(\lambda)}\\
&& +R_{m,m}(\lambda,t) ,\ \ \ m=1,2, \\
{} [\rhobar_t]_{1,2} &=& {\textstyle \frac 12} \e^{\i t\varepsilon_{-\Delta}(\lambda)} +R_{1,2}(\lambda,t),\\
{} [\rhobar_t]_{2,1} &=& {\textstyle \frac 12} \e^{\i t\varepsilon_{\Delta}(\lambda)}+R_{2,1}(\lambda,t),
\end{eqnarray*}
where the numbers $\varepsilon$ are given in \fer{i4}. The remainder terms satisfy $|R_{m,n}(\lambda,t)|\leq C\lambda^2$, uniformly in $t\geq 0$, and they can be decomposed into a sum of a constant part (in $t$) and a decaying one,
$
R_{m,n}(\lambda,t) = \av{\av{p_{n,m}}}_\infty-\delta_{m,n} \frac{\e^{-\beta E_m}}{Z_{\s,\beta}} +R'_{m,n}(\lambda,t),
$
where $|R'_{m,n}(\lambda,t)|=O(\lambda^2\e^{-\gamma t})$, with $\gamma=\min\{{\rm Im}\varepsilon_0, {\rm Im}\varepsilon_{\pm\Delta}\}$. Therefore, to second order in $\lambda$, convergence of the populations to the
equilibrium values (Gibbs law), and decoherence occur exponentially
fast, with rates $\tau_T=[\mathrm{Im}\varepsilon_0(\lambda)]^{-1}$
and $\tau_D=[\mathrm{Im}\varepsilon_\Delta(\lambda)]^{-1}$,
respectively. In
particular, coherence of the initial state stays preserved on time
scales of the order $\lambda^{-2}[|c|^2
\xi(\Delta)+(b-a)^2\xi(0)]^{-1}$, c.f. \fer{i4}.

\section{\label{sect:discussion} Discussion}

Relation \fer{introdd1} gives a detailed picture of the dynamics of
averages of observables. The resonance energies $\varepsilon$ and
the functionals $R_\varepsilon$ can be calculated for concrete
models, to arbitrary precision (in the sense of rigorous perturbation theory
in $\lambda$). See \fer{i1}-\fer{i4} for explicit expressions for
the qubit, and the illustration above for an initially coherent
superposition given by \fer{illust}. In the present work we use relation
\fer{introdd1} to discuss the processes of thermalization and
decoherence of a qubit. In \cite{MSB2} we present, besides a proof of \fer{introdd1}, applications to energy-preserving (non-demolition) interactions and to registers of arbitrarily many qubits. It would be interesting to apply the techniques
developed here to the analysis of the transition from quantum to classical behaviour (see \cite{BBBD,DBV}).

In the absence of interaction
($\lambda=0$) we have $\varepsilon=e\in{\mathbb R}$, see \fer{int1}.
Depending on the interaction each resonance energy $\varepsilon$ may
migrate into the upper complex plane, or it may stay on the real
axis, as $\lambda\neq 0$. The averages $\av{A}_t$ approach their
ergodic means $\av{\av{A}}_\infty$ if and only if $\mathrm{Im}
\varepsilon
>0$ for all $\varepsilon\neq 0$. In this case the convergence takes place
on the time scale $[\mathrm{Im}\varepsilon]^{-1}$. Otherwise
$\av{A}_t$ oscillates.
A sufficient condition for decay is that ${\rm Im}\delta_e^{(s)} <0$
(and $\lambda$ small, see \fer{int1}).

There are two kinds of processes which drive the decay:
energy-exchange processes and energy preserving ones. The former are
induced by interactions enabling 
processes of absorption and emission of field quanta with energies
corresponding to the Bohr frequencies of $\s$ (this is the ``Fermi
Golden Rule Condition'' \cite{BFS,FM1,Mlso,MMS1,MMS2}). Energy
preserving interactions suppress such processes, allowing only for a
phase change of the system during the evolution (``phase damping'',
\cite{PSE,BBD,DBV,DG,JZKGKS,MP,SGC}).

Even if the initial density matrix, $\rho_{t=0}$, is a product of the system and reservoir density
matrices, the density matrix, $\rho_t$, at any subsequent moment of
time $t>0$ is not of the product form. The evolution
creates the system-reservoir entanglement. We develop a formula for $\av{A}_t-\av{\av{A}}_\infty$ for all 
observables $A$ of any $N$-level system $\s$ in \cite{MSB2}.
If the system has the property of return to equilibrium, i.e.,
if $\xi(\Delta)>0$, then
$
[\rhobar_\infty]_{m,n} = \delta_{m,n}\frac{e^{-\beta E_m}}{\tr_\s(\e^{-\beta H_\s})} + O(\lambda^2)$. Hence the Gibbs distribution is obtained by first letting $t\rightarrow\infty$ and then $\lambda\rightarrow 0$. A similar observation in the setting of the quantum Langevin equation has been made in \cite{BK}. 
If $\rho_0$ is an arbitrary initial density matrix on
$\h_\s\otimes\h_\r$ then our method yields a similar result, see \cite{MSB2}.

Equations \fer{i1}, \fer{i2} and \fer{i4} define the decoherence time
scale,  $\tau_D=[\mathrm{Im}\varepsilon_\Delta(\lambda)]^{-1}$, and
the thermalization time scale,  $\tau_T=[\mathrm{Im}\varepsilon_0(\lambda)]^{-1}$. We should compare
$\tau_D$ with the
decoherence time scales and with computational time
scales in real systems. The former vary from $10^4$s for nuclear spins in
paramagnetic atoms to $10^{-12}$s for electron-hole excitations in
bulk semiconductors (see e.g. \cite{DiV}).

In the ubiquitous spin-boson model \cite{LCDFGZ}, obtained as a two-state truncation of a double-well system or an atom interacting with a Bose field, the Hamiltonian is given by \fer{ii1} with $H_\s = -\frac 12\hbar\Delta_0\sigma_x +\frac 12\epsilon\sigma_z$ and $v=\sigma_z\otimes\varphi(g)$. Here, $\sigma_x$, $\sigma_z$ are Pauli spin matrices, $\epsilon$ is the ``bias'' of the asymmetric double well, and $\Delta_0$ is the ``bare tunneling matrix element''. In the canonical basis, whose vectors represent the states of the system localized in the left and the right well, $H_\s$ has the representation
\begin{equation}
H_\s = \frac 12
\left[
\begin{array}{cc}
\epsilon & -\hbar\Delta_0 \\
-\hbar\Delta_0 & -\epsilon
\end{array}
\right].
\label{sb1}
\end{equation}
The diagonalization of $H_\s$ yields $H_\s\cong {\rm diag}(E_+,E_-)$, where $E_\pm = \pm\frac 12\sqrt{\epsilon^2+\hbar^2\Delta_0^2}$. The operator $v=\sigma_z\otimes \varphi(g)$ is represented in the basis diagonalizing $H_\s$ as \fer{michael*}, with $a=-b= -(\frac{\hbar^2\Delta_0^2}{\epsilon^2}+1)^{-1/2}$ and $c=\frac 12 (\frac{\epsilon^2}{\hbar^2\Delta_0^2}+1)^{-1/2}$.


\end{document}